\documentclass [aps,prd,eqsecnum,showpacs,superscriptaddress,twocolumn,floatfix,preprintnumbers,altaffilletter]{revtex4}

\usepackage{graphicx}
\usepackage{amssymb}
\usepackage{amsmath}
\usepackage{amsfonts}

\begin{document}

\title[Detecting GW Transients Associated with SGR QPOs]{Detecting Long-Duration Narrow-Band Gravitational Wave Transients Associated with Soft Gamma Repeater Quasi-Periodic Oscillations}

\author{David~Murphy}
\email{djm2131@columbia.edu}
\affiliation{Department of Physics, Columbia University, New York, NY 10027, USA}

\author{Maggie~Tse}
\affiliation{Department of Physics, Columbia University, New York, NY 10027, USA}

\author{Peter~Raffai}
\affiliation{Department of Physics, Columbia University, New York, NY 10027, USA}
\affiliation{MTA-ELTE EIRSA ''Lendulet'' Astrophysics Research Group}

\author{Imre~Bartos}
\affiliation{Columbia University, Department of Physics, New York, NY 10027, USA}

\author{Rubab~Khan}
\affiliation{Department of Astronomy, Ohio State University, Columbus, OH 43210, USA}

\author{Zsuzsa~M\'{a}rka}
\affiliation{Columbia University, Department of Physics, New York, NY 10027, USA}

\author{Luca~Matone}
\affiliation{Columbia University, Department of Physics, New York, NY 10027, USA}

\author{Keith~Redwine}
\affiliation{Department of Physics and Astronomy, Johns Hopkins University, Baltimore, MD 21218, USA}

\author{Szabolcs~M\'{a}rka}
\affiliation{Columbia University, Department of Physics, New York, NY 10027, USA}

\begin{abstract}
We have performed an in-depth concept study of a gravitational wave data analysis method which targets repeated long quasi-monochromatic transients (triggers) from cosmic sources. The algorithm concept can be applied to multi-trigger data sets in which the detector-source orientation and the statistical properties of the data stream change with time, and does not require the assumption that the data is Gaussian. Reconstructing or limiting the energetics of potential gravitational wave emissions associated with quasi-periodic oscillations (QPOs) observed in the X-ray lightcurve tails of soft gamma repeater flares might be an interesting endeavour of the future. Therefore we chose this in a simplified form to illustrate the flow, capabilities, and performance of the method. We investigate performance aspects of a multi-trigger based data analysis approach by using $\mathcal{O} \left( 100 \, \text{s} \right)$ long stretches of mock data in coincidence with the times of observed QPOs, and by using the known sky location of the source. We analytically derive the PDF of the background distribution and compare to the results obtained by applying the concept to simulated Gaussian noise, as well as off-source playground data collected by the 4-km Hanford detector (H1) during LIGO's fifth science run (S5). We show that the transient glitch rejection and adaptive differential energy comparison methods we apply succeed in rejecting outliers in the S5 background data. Finally, we discuss how to extend the method to a network containing multiple detectors, and as an example, tune the method to maximize sensitivity to SGR 1806$-$20 flare times.
\end{abstract}

\maketitle

\section{Introduction} \label{sec: intro}

Soft gamma repeaters (SGRs) are mostly galactic objects that emit short bursts of X-ray and gamma radiation at irregular intervals. At times these sources emit exceptionally energetic flares that last hundreds of seconds, with peak luminosities of up to $10^{46}\,$erg s$^{-1}$ \cite{WS}. The most luminous SGR flare observed to date was recorded on 27 December 2004, and originated from SGR 1806$-$20 \cite{Is}. Quasi-periodic oscillations (QPOs) were observed at late times in the lightcurve tail by the Rossi X-Ray Timing Explorer (RXTE) \cite{RXTEReview} and the Ramaty High Energy Solar Spectroscopic Imager (RHESSI) \cite{RHESSIReview} satellites, most notably at $92.5\,$Hz and $626.5\,$Hz \cite{WS,Is}. Although the exact timing of the QPO signals is unclear since the peak flare saturated both detectors, the signals can be unambiguously observed beginning approximately $170-200\,$s after the initial flare \cite{WS,Is}.

The $\sim 8\,$s period of the lightcurve tail observed during the December 2004 hyperflare \cite{Is}, as well as the phase-dependence of the QPO features, suggests that the parent object might be a type of slowly rotating neutron star. In this model the phase dependence can be interpreted as localization to a specific region of the star's surface, and the observed frequencies may be related to particular seismic modes \cite{WS,Is}, making QPOs an interesting probe of the astroseismology of SGRs. In particular, if QPOs are indeed related to seismic vibrations of a neutron star, then we can estimate the thickness of the star's crust, and possibly infer what type of material makes up the star's interior \cite{Watts07}. It has also been suggested that changes in the observed QPOs over time may be indicative of physical changes in the source, such as a decaying magnetic field \cite{EM}.

The magnetar model \cite{MagnetarModel} proposes that SGRs are a type of highly magnetized NS with a surface magnetic field in excess of $10^{15}\,$G. In this model QPOs can be explained by a starquake event, in which the star's magnetic field becomes twisted and then violently rearranges, exciting crustal or global seismic modes and releasing gamma ray bursts with the observed QPO structures \cite{Sc, Pa}. Numerical studies in which the magnetar is treated as a coupled system consisting of an elastic crust and a magnetohydrodynamic fluid core have demonstrated a number of interesting effects. These include the dynamical generation of instabilities and asymmetric quasi-equilibria in the magnetic field \cite{Lasky}, as well as a coupling between discrete crustal modes and continuous bands of global Alfv\'{e}n modes, with QPOs appearing in the late time starquake oscillations near the edges of the Alfv\'{e}n spectral bands \cite{Levin, LevinI, LevinII, Sotani, EoM}. This numerical work also suggests that some QPOs may be exceptionally strong because they are close in frequency to one of the crustal modes \cite{LevinI, LevinII, EoM}. The same dynamics may also drive the emission of gravitational waves (GWs) by exciting the magnetar's seismic modes; in particular, crustal f-modes \cite{GWsourcereview}. Although it is argued that at realistic values of the magnetic field strength f-modes are unlikely to be sufficiently excited to produce an observable signal in second-generation gravitational wave detectors \cite{LevinGW,Zink}, Zink et al. note that low-frequency modes up to 100 Hz remain unstudied, and may be sufficiently excited to produce detectable GW signals \cite{Zink}.

A number of searches for gravitational radiation associated with SGRs and magnetars have been conducted. In particular, a search for GWs associated with f-mode ringdowns in six magnetars targeting the frequency ranges $1-3\,$kHz, $100-200\,$Hz, and $100-1000\,$Hz, with time windows of $\sim 200\,$ms, found no evidence of a GW signal \cite{6Magnetars}. Other searches targeting short-duration ($\sim 0.3\,$s) GWs from SGRs \cite{GWSGR} as well as stacked GW emissions associated with the SGR 1900$+$14 storm event \cite{SGR1900storm} also found no evidence of GWs associated with SGR bursts at energies accessible by past interferometric gravitational wave detectors.

Our work extends a search algorithm developed by Matone and M\'{a}rka \cite{OldMethods}, targeting narrow-band, long-duration GW signals associated with SGR QPOs. This algorithm is based on an excess energy method and thus does not presume any particular waveform for the emitted GW signal. In addition, the algorithm also does not require exact knowledge of the timing of the GW signal relative to the electromagnetic trigger; we choose conservative $\mathcal{O} \left( 100 \, \text{s} \right)$ long time windows encompassing the entire flare event. The algorithm was previously applied to LIGO data recorded during the December 2004 SGR 1806$-$20 hyperflare event to constrain the energetics of any associated GW signal, and set a lowest bound of $7.67 \times 10^{46}\,$erg on the total GW emissions associated with the QPO at $92.5\,$Hz \cite{hyperflare}. A similar algorithm targeting long-duration GW transients using a complementary cross-power technique is discussed in \cite{STAMP}, and is also relevant to searches for GWs from SGRs in the LIGO data.

The LIGO network of interferometric gravitational wave detectors has completed 6 science runs to date. Science runs 5 and 6 lasted from November 4th, 2005 to September 30th, 2007, and from July 7th, 2009 to October 20th, 2010, respectively. During S5 alone $\sim 150$ EM flares originating from SGR 1806$-$20 were observed. If SGR flares are indeed coupled to seismic vibrations of a neutron star as suggested by the magnetar model, one might expect that each burst should excite the same, characteristic frequencies of the star's crust regardless of the luminosity of the flare, leading to repeated GW emissions at the same frequencies. As a result, a comprehensive algorithm analyzing all triggers and QPO frequencies overlapping with the LIGO data set for evidence of repeated GW emissions from SGR flares might prove itself more sensitive than an algorithm targeting a single burst or QPO frequency. Although we illustrate the concept through a $10\,$Hz frequency band centered on the $92.5\,$Hz QPO observed during the December 2004 SGR 1806$-$20 hyperflare, the algorithm can be applied to any frequency window of interest, allowing the method to target each of the observed QPO frequencies.

The algorithm concept presented has been designed to efficiently process large, multi-detector data sets containing multiple SGR flares from a single source, and includes a number of data quality measures intended to account for nonstationarity and deviations from Gaussianity in the data stream of a physical detector operating for months or years. The method, like the methods mentioned above, benefits from knowledge of the arrival time of an electromagnetic counterpart to the hypothesized GW signal, as well as knowledge of the sky position of the source. In addition, we further assume that the recurrent SGR bursts contain the same QPO structures and seismic modes observed in hyperflares, and introduce a new detection criterion based on the two-sample Kolmogorov-Smirnov (K-S) test and Fisher's method, allowing us to increase the sensitivity of the algorithm relative to analyzing a single event. We also discuss the problem of variable detector orientation for each trigger, and discuss how to optimize the concept accordingly. We show that with the proper data conditioning procedures the behaviour of real data can be modeled well mathematically and approximated numerically through gaussian statistics.

The paper is organized as follows: in Sec.~\ref{sec: theory} we discuss the algorithm and analytically derive the PDF of the background distribution for Gaussian white noise. In Sec.~\ref{sec: pipeline} we discuss the implementation of the algorithm as a concept pipeline and data quality measures. Sec.~\ref{sec: sim} verifies the implementation by applying the pipeline to randomly generated Gaussian noise and comparing the results to the predictions of the mathematical model. Finally, in Sec.~\ref{sec: data} we present a concept study of the off-source detector background and orientation using playground data segments from LIGO's fifth science run \cite{LIGO}.

\section{Mathematical Model of the Algorithm} \label{sec: theory}

The time domain algorithm splits the data stream by bandpass filtering into three channels of a specific bandwidth motivated by X-ray observation constraints $\Delta f$: the QPO channel, which is centered on the QPO frequency, and the $up$ and $down$ channels, centered at $f_\text{up} = f_\text{QPO} + \Delta f$ and $f_\text{down} = f_\text{QPO} - \Delta f$, respectively. The bandwidth $\Delta f$ for example can be derived from the measured full width at half maximum (FWHM) of the QPOs observed in the EM (X-ray) emissions \footnote{For the case of the December 2004 SGR 1806$-$20 hyperflare a table summarizing the frequencies and FWHMs of the observed QPOs can be found in \cite{hyperflare}. If assuming, based on the magnetar model, that the QPO frequencies and bandwidths are the same as the f-modes of the neutron star, these can also be the frequencies and bandwidths of the gravitational radiation \cite{LevinGW,Zink}.}. We define the energy of a strain time series $\xi (t)$ in an interval $\Delta t$ to be
\begin{equation} \label{energy}
E = \int_{\Delta t} dt \left| \xi \left( t \right) \right|^2.
\end{equation}
We can compare the energy contained in the QPO band to the energy contained in the adjacent bands during an interval $\Delta t$ by computing the excess energy
\begin{equation} \label{Eexcess}
E^{\text{excess}} = E_{\text{QPO}} - w \left( E_{\text{up}} + E_{\text{down}} \right),
\end{equation}
where $w$ is a locally determined averaging factor \footnote{Naively one would choose $w = 1/2$, but the noise floor of a physical instrument is neither flat, linear, or constant across all frequency bands. Instead, we use a locally determined value that is discussed in more detail in Sec.~\ref{subsec: DQ}.}. Defining the excess energy in this manner allows us to isolate the contribution of a GW signal associated with the QPO to the total energy of the data stream while simultaneously suppressing broadband instrumental noise. It is also necessary to account for temporal non-stationarity in the data stream since we perform a comprehensive analysis with data spanning several years, during which changes in the properties of the detector noise are too significant to be ignored. We accomplish this by computing the excess energy in the time interval $\left( t_\text{center} \hspace{0.5mm} , \hspace{0.5mm} t_\text{center} + \Delta t \right)$, the excess energy in the time interval $\left( t_\text{center} - \Delta t \hspace{0.5mm} , \hspace{0.5mm} t_\text{center} \right)$, and then computing the difference in excess energy
\begin{align} \label{dEexcess}
\Delta E^{\text{excess}}&=& \\
&=& E^{\text{excess}}_{\left( t_\text{center} \hspace{0.5mm} , \hspace{0.5mm} t_\text{center} \hspace{0.25mm} + \hspace{0.25mm} \Delta t \right)} - E^{\text{excess}}_{\left( t_\text{center} \hspace{0.25mm} - \hspace{0.25mm} \Delta t \hspace{0.5mm} , \hspace{0.5mm} t_\text{center} \right)}\nonumber
\end{align}
resulting in a quantity which can be compared directly across different times. For on-source measurements we choose $t_\text{center}$ to be the time of an observed trigger \footnote{We define the trigger time to be the arrival time of the observed EM burst.}, and for off-source measurements $t_\text{center}$ is randomly sampled from stretches of data disjoint from the region hypothesized to contain a GW signal. The value of $\Delta t$ is chosen based on the timescale of the SGR flares, and is the same for both on-source and off-source measurements: since we do not, in general, have detailed information regarding the length of every SGR burst, we choose $\Delta t$ to be the full duration of the longest observed lightcurve tail from a given SGR as an illustration here. For SGR 1806$-$20 this is $\Delta t = 300\,$s based on the December 2004 hyperflare. This choice of $\Delta t$ is intended to be conservative to ensure that we capture associated GW signals. The method for calculating $\Delta E^{\text{excess}}$ is depicted schematically in Fig.~\ref{fig: w_schematic}.

By considering each sampling point in the data stream as an independent, identically distributed normal random variable we can construct a theoretical model for the algorithm. In the following section we derive an explicit formula for the probability density function of $\Delta E^{\text{excess}}$. These results are then used in Sec.~\ref{sec: sim} to check the algorithm's performance by comparing numerical simulations to the theoretical results.

\subsection{Sampling Distribution of $\Delta E^{\text{excess}}$}

Here we model the strain noise of the detector by band-limited Gaussian white noise for which each sampling point, $\xi_i$, is an independent \footnote{For a band-limited noise sample with bandwidth $\Delta f$ and length $\Delta t$, the independence of the sampling points is an approximation that is only valid when $\Delta t\gg \Delta f^{-1}$.} Gaussian random variable with zero mean and standard deviation $\sigma$. The energy of the strain noise in a time interval $\Delta t$ is given by the discrete analogue of \eqref{energy}
\begin{equation} \label{Strain_E}
E = \frac{{1}}{{F_\text{s}}} \sum\limits_{i = 1}^n {\xi_i}^2,
\end{equation}
where $F_\text{s}$ is the sampling frequency and $n = F_\text{s} \Delta t$ is the total number of sampling points. $E$ follows a $\chi^2$-distribution with probability density function (PDF) \cite{MathMethods}
\begin{equation} \label{E_single_channel}
f_{E} \left( x \right) = \begin{cases}\frac{{F_\text{s}}^{n/2}}{{\sigma^n 2^{n/2} \Gamma \left( n/2 \right) }} x^{n/2 - 1} e^{- \frac{{F_\text{s}}}{{2 \sigma^2}} x}&\text{for $x \ge 0$}\\0 &\text{for $x < 0$}\end{cases}
\end{equation}
Assuming that the channels are uncorrelated and the signal energy is distributed isotropically in frequency space it suffices to choose $w = 1/2$ in \eqref{Eexcess}, and we can then define a random variable $\Delta E^{\text{excess}}$ as
\begin{align}
\Delta E^{{\text{excess}}} = E_{{\text{after\hspace{1mm}flare}}}^{{\text{excess}}} - E_{{\text{before\hspace{1mm}flare}}}^{{\text{excess}}} = \\
= \left[ E_\text{QPO}^\text{after} - \frac{1}{2} \left( E_\text{up}^\text{after} - E_\text{down}^\text{after} \right) \right] - \nonumber \\
- \left[ E_\text{QPO}^\text{before} - \frac{1}{2} \left( E_\text{up}^\text{before} - E_\text{down}^\text{before} \right) \right] \nonumber
\end{align}
where each of the terms has probability density function \eqref{E_single_channel}. The PDF of $\Delta E^{\text{excess}}$ can be found using an identity for characteristic functions (here denoted by $\varphi$) \cite{ProbTheory}
\begin{equation} \label{pdftheorem}
\varphi _{\sum\limits_i {a_i X_i } } \left( s \right) = \prod\limits_i {\varphi _{X_i } } \left( {a_i s} \right).
\end{equation}
The characteristic function associated with \eqref{E_single_channel} is
\begin{equation}
\varphi _{E} \left( s \right) = \frac{1}{{\left( {1 - \frac{{2\sigma ^2 }}{{Fs}}is} \right)^{n/2} }} \hspace{2mm},
\end{equation}
so, using \eqref{pdftheorem} and the inverse Fourier transform we find that
\begin{align} \label{pdf_dEexcess}
f_{\Delta E^{{\text{excess}}} }& \left( x \right) = \\
&= \frac{1}
{{2\pi }}\int\limits_{ - \infty }^\infty ds \hspace{1mm} {\frac{{\cos \left( {s x} \right)}}
{{\left( {1 + \frac{{4\sigma ^4 }}
{{F_s^2 }}s^2 } \right)^{n/2} \left( {1 + \frac{{\sigma ^4 }}
{{F_s^2 }}s^2 } \right)^n }}} \hspace{2mm} \nonumber.
\end{align}
The cumulative distribution function (CDF) is
\begin{align} \label{cdf_dEexcess}
F&_{\Delta E^{{\text{excess}}} } \left( x \right) = \\
&= \frac{1}
{{2\pi }} \int\limits_{- \infty}^x du \int\limits_{ - \infty }^\infty ds \hspace{1mm} {\frac{{\cos \left( {su} \right)}}
{{\left( {1 + \frac{{4\sigma ^4 }}
{{F_s^2 }}s^2 } \right)^{n/2} \left( {1 + \frac{{\sigma ^4 }}
{{F_s^2 }}s^2 } \right)^n }}} \hspace{2mm} \nonumber.
\end{align}
Note that the CDF can similarly be calculated for any other values of $w$.

\section{Description of the Pipeline} \label{sec: pipeline}

\begin{figure}[h]
        \vspace{11mm}
	\centering
	\includegraphics[width=3.4in]{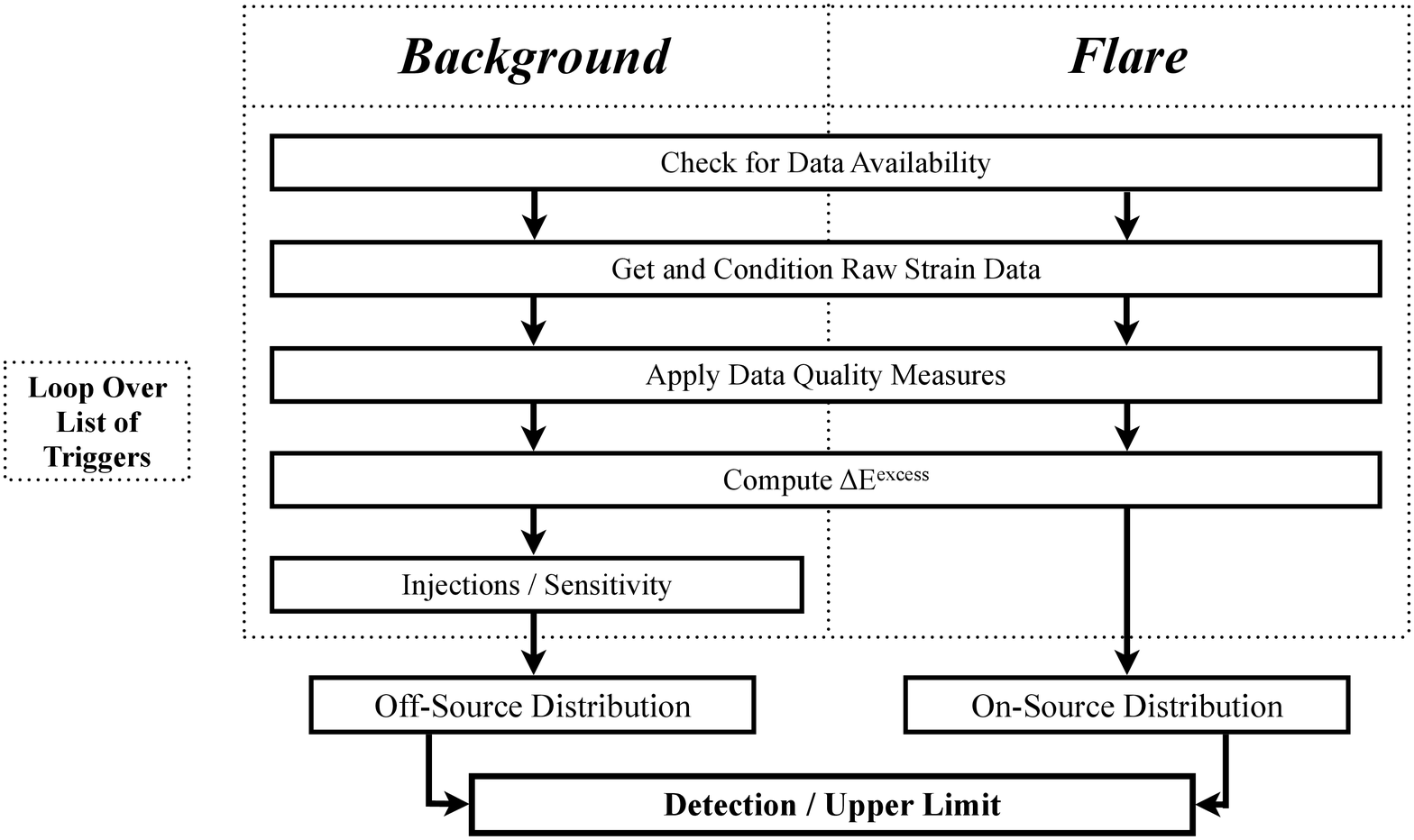}
	\caption{A block diagram illustrating the steps in the analysis pipeline concept. The data conditioning step is discussed in Sec.~\ref{subsec: conditioning}. Data quality measures are intended to account for variability in the quality of data recorded by a physical instrument and are discussed in Sec.~\ref{subsec: DQ}. The injection step is optional, and is used to place an upper limit in the event that the pipeline fails to make a detection. \label{fig: flowchart}}
\end{figure}

The pipeline is designed to sequentially process the LIGO data stream in the time domain and take $\Delta E^\text{excess}$ measurements around times associated with SGR flares. We distinguish between flare measurements and background measurements. Flare measurements are performed at the time of an observed trigger, whereas background measurements are performed at times before or after a trigger and used to characterize the detector background. Fig.~\ref{fig: flowchart} depicts a schematic of the algorithm.

Using a list of the start and end times of detector lock stretches that contain one or more astrophysical triggers, the pipeline sequentially processes each trigger, performing a single measurement at the time of the trigger, and a fixed number of background measurements before and after the trigger. For each measurement the data is filtered to isolate the QPO band and the two adjacent frequency bands. The algorithm then applies data quality measures to the filtered data streams as described in Sec.~\ref{subsec: DQ}, and computes $\Delta E^\text{excess}$ as defined by \eqref{dEexcess}. After iterating over the full list of triggers the resulting background and flare distributions are compared using the two-sample K-S test \cite{KStest}, and detection is claimed if the distributions are distinguishable with $5 \sigma$ confidence. Otherwise, we inject a simulated waveform of known energy into a single pre-flare background measurement corresponding to each trigger to determine the sensitivity --- defined to be the minimum injected energy such that the background and foreground distributions meet our detection criteria --- and set an upper limit on the energy of any GW signal contained in the data stream.

\subsection{Data Conditioning} \label{subsec: conditioning}

The ingoing data stream is conditioned by applying a series of three zero-phase, bandpass, Butterworth filters, resulting in the upper, lower, and QPO channels. Sufficiently large buffer time intervals are included and cut off after filtering at both ends of the data stream to make sure that no transients remain in the data due to filtering. The difference in excess energy is then calculated according to \eqref{dEexcess}. Injections are performed by generating the plus and cross components ($h_+ \left( t \right)$ and $h_\times \left( t \right)$, respectively) of the simulated GW signal with a given root-sum-square strain ($h_\text{rss}$)
\begin{equation}
h_\text{rss} \equiv \sqrt{ \int dt \left ( h^2_+ \left( t \right) + h^2_\times \left( t \right) \right)  }
\end{equation}
and adding it to the input data prior to the data conditioning step. We can write
\begin{equation} \label{eqn: inj}
h \left( t \right) = F_+ \cdot h_+ \left( t \right) + F_\times \cdot h_\times \left( t \right)
\end{equation}
where and $F_+$ and $F_\times$ are antenna factors determined by the orientation of the detector relative to the source at the time of a given trigger and the choice of the GW polarization frame. The antenna factors satisfy $0 \le F_+,F_\times \le 1$ and encode the directional sensitivity of the detector to an incoming plus- ($F_+$) or cross- ($F_\times$) polarized GW \cite{DPF}. The injection process is performed once per trigger, resulting in a simulated foreground distribution that can be used to determine the sensitivity of the algorithm.

The root-sum-square strain measured at the detector ($h_\text{rss}$) can be translated into a total physical energy ($E_\text{GW}^\text{iso}$) that the source would emit in GWs if the emission was isotropic \cite{hyperflare}:
\begin{equation} \label{eqn: E_GW}
E_\text{GW}^\text{iso} = \frac{\pi^2c^3}{G} d^2 f_0^2 h_\text{rss}^2 ,
\end{equation}
where $d$ is the distance between the detector and the source, and $f_0$ is the central frequency of the GW signal \footnote{The isotropic equivalent energy of the GW emission, $E_\text{GW}^\text{iso}$, is measured in $erg$s, and should not be confused with the energy of a strain time series defined in \eqref{energy} and measured in seconds.}. In the event of a non-detection this relation allows us to use the measured sensitivity to constrain the energetics of GW emissions at the source.

\subsection{Data Quality} \label{subsec: DQ}

A number of measures have been implemented to account for instrumental transients and the variability in data quality across long stretches of time. We rely primarily on three techniques: (i) the application of data quality flags, (ii) transient glitch rejection, and (iii) the use of a locally-determined weighting factor $w$ to measure the excess energy (defined by \eqref{Eexcess}).

\begin{enumerate}
\item Data quality flags are constructed by the LSC's detector characterization working group to characterize noise contamination in the detector's data stream, and mark the times of known environmental noise sources and detector calibration issues. The flags are divided into categories 1, 2, 3, and 4 based on severity and duration. Here we reject any data marked with a flag designated category 1 --- times when the detector was not taking data in the design configuration --- or category 2 --- times when well-understood instrumental glitches were observed. We also avoid standard and significant line features while selecting the frequency bands, and we propose to apply lock-in filtering techniques to mitigate the effects of any potentially unavoidable line features. Detailed descriptions of the flags can be found in \cite{DQFlags}.

\item Transient glitch rejection provides a second method for reducing noise contamination. Since we expect a long-duration signal lasting hundreds of seconds we can safely reject transient glitches over much smaller time scales. This is accomplished using a method similar to the glitch veto algorithm discussed in \cite{hyperflare}: we divide the data stream of each, filtered channel into tiles of length $\Delta \tau$, typically of order $1\,$s or less, compute distributions of the energy per tile for each of the three channels, and recursively veto any tiles with energy more than $\pm 4$ standard deviations from the mean tile energy. We then divide each channel by a duty factor
\begin{equation} \label{DF}
DF = 1 - \frac{{\text{length} \left( \text{veto} \right)}}{{\text{length} \left( \text{data} \right)}}
\end{equation}
to renormalize the remaining data in this example \footnote{In general, when a nontrivial veto step is applied to the data stream of a physical detector, independent measurements will have different lengths of data removed by the veto based on the magnitude and duration of any glitches. The duty factor is proportional to the length of data remaining, and is necessary to renormalize the post-veto data stream so that independent measurements have comparable total energies. If the signal exhibits strong time dependence as it evolves this renormalization will introduce model/signal dependent distortions to the signal energy measurement, nevertheless it is efficient for the locally flat background.}. The veto is applied to the strain time series of each filtered channel independently since glitches are not generally distributed evenly in frequency space. This process is iterated until either all outliers have been removed or the fraction of data vetoed passes a cutoff value, typically chosen to be 10\%, at which point the data segment is rejected as unreliable.

\item Another source of data contamination arises from the curvature of the LIGO noise floor. Naively choosing $w = 1/2$ in the definition of the excess energy statistic \eqref{Eexcess} results in a quantity that does not properly account for differences in the background noise across multiple frequency bands. Since the properties of the detector noise change with time we are forced to calculate an appropriate value of $w$ for each measurement \footnote{This nonstationarity can be seen in Fig.~\ref{fig: S5_SF_plot}, which shows a plot of $w$ calculated using LIGO data.}. With this in mind we define a local weighting factor $w$ by averaging the four quantities
\begin{equation} \label{w}
w_{ \pm 1, \pm 2 } = \left| \frac{{E_\text{QPO}}}{{E_\text{up} + E_\text{down}}} \right| _{t_w = \pm t_{w,1}, \hspace{1mm} \pm t_{w,2}}
\end{equation}
calculated by summing the energies of each channel in the time interval $[ t_w - \Delta t , t_w + \Delta t]$ for each of the four times $t_w = \pm t_{w,1},  \pm t_{w,2}$. This is depicted schematically for a hypothetical measurement in Fig.~\ref{fig: w_schematic}. \\
\begin{figure}[h]
\centering
\includegraphics[width=3.4in]{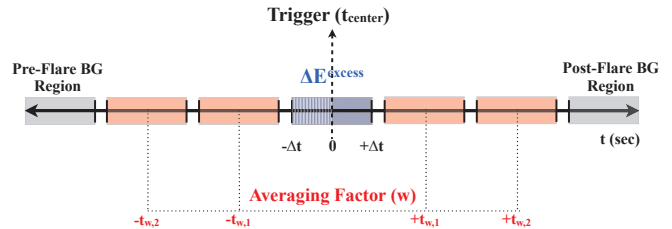}
\caption{A schematic illustrating the regions of the time-series used to perform a $\Delta E^\text{excess}$ measurement with a locally determined weighting factor $w$. The measurement is centered about a time $t_\text{center}$ which is either the time of an observed SGR trigger or randomly sampled from the background data. The value of $w$ is obtained by calculating the ratio \eqref{w} using the data contained in the interval $\left[ t_w - \Delta t, t_w + \Delta t \right]$ for each of the four times $t_w = \pm t_{w,1},  \pm t_{w,2}$. These four values are averaged to estimate $w$, and then $\Delta E^\text{excess}$ is computed using this value in \eqref{Eexcess} and \eqref{dEexcess}. The part of the diagram corresponding to $\Delta E^\text{excess}$ is divided into a solid region and a dashed region to emphasize that it is the difference in excess energy between these two regions which we ultimately calculate using the locally determined $w$. \label{fig: w_schematic}}
\end{figure} \\
We choose $\Delta t$ to be the same time interval used for computing excess energy measurements, and $t_{w,1}$ and $t_{w,2}$ to be the minimum adjacent times which ensure that none of the measurements overlap. With this scheme we can use the data immediately preceding and following the region used for a given $\Delta E^\text{excess}$ measurement to determine the appropriate value of $w$ in \eqref{Eexcess}. Defined in this way, $w$ accounts for differences between channels due to the shape of the noise floor, as well as changes in the noise floor over time.
\end{enumerate}

\section{Gaussian Noise Simulations} \label{sec: sim}

In order to validate the concept and test the mathematical model of section \ref{sec: theory} we simulate the result of applying the pipeline to a set of detected SGR 1806$-$20 flares coinciding with the data from LIGO's fifth science run (S5). This is accomplished by configuring the pipeline to perform a realistic study but replacing the LIGO data stream with stretches of randomly generated Gaussian white noise. This is not unreasonable as the data quality cut studies indicated that the ''cleaned'' data is very similar to a Gaussian random set.

\subsection{Simulated Detector Background} \label{sec: det_bg}

We construct a background sampling distribution by calculating $\Delta E^{\text{excess}}$ for simulated data consisting of 240 seconds of randomly generated Gaussian white noise with strain spectral noise density $\tilde{n} = 10^{-22}\,$Hz$^{-1/2}$. The value of $\tilde{n}$ is chosen so that the simulated noise is similar in magnitude to the LIGO noise floor during S5. We choose $f_\text{QPO} = 92.5\,$Hz and $\Delta f = 10\,$Hz, corresponding to the frequency and FWHM of the strongest QPO signal observed in the lightcurve tail of the 27 December, 2004 SGR 1806$-$20 hyperflare event, and $\Delta t = 30\,$s \cite{hyperflare}. This value of $\Delta t$ is shorter than the $\sim 300\,$s duration of the lightcurve tail observed during the SGR 1806$-$20 hyperflare, nevertheless for the purposes of a concept study, is less computationally expensive while providing essentially the same insight. We also compute the theoretical PDF of the background distribution using the same parameters to compare the theoretical result \eqref{pdf_dEexcess} to the simulation.

\begin{figure}
\vspace{9mm}
\centering
\includegraphics[width=3.4in]{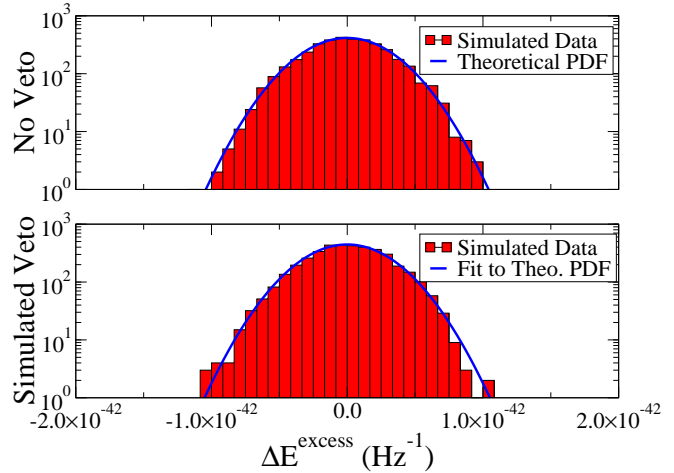}
\caption{Detector background with and without a simulated data quality veto step. The parameters used to generate the Gaussian noise are $n = 491520$ and $\tilde n = 10^{-22}\,$Hz$^{-1/2}$, where $n = F_s \Delta t$ is the total number of sampling points used to make a $\Delta E^\text{excess}$ measurement and $\tilde{n}$ is the strain spectral density of the noise. The upper plot shows the agreement between the detector background with no vetoes (histogram) and the theoretical PDF (solid curve) obtained by numerically integrating \eqref{pdf_dEexcess}. The lower plot shows a fit of the theoretical PDF to the modified detector background after incorporating simulated data quality vetoes. The parameters of the fit, obtained using nonlinear least-squares regression to fit the theoretical CDF \eqref{cdf_dEexcess} to the empirical CDF of the resulting detector background, are $n = 493285$ and $\tilde{n} = 10^{-22}\,$Hz$^{-1/2}$. \label{fig: GWN_sim}}
\end{figure}

In anticipation of later work utilizing LIGO data, we also construct a second background distribution incorporating a simulated veto step to demonstrate that our algorithm is robust with respect to the inclusion of data quality vetoes. This is accomplished by randomly removing 5\% of the sampling points from the strain time series of each, filtered channel independently, and then applying a duty factor \eqref{DF} to each channel. We also obtain a fitted PDF by using nonlinear least squares regression to fit the theoretical CDF \eqref{cdf_dEexcess} to the empirical CDF of the resulting background distribution. Fig.~\ref{fig: GWN_sim} depicts the results of these simulations. \\
In order to characterize the goodness of fit we compute the mean percentage error (MPE)
\begin{equation}
\text{MPE} = \frac{1}{n} \sum_{i=1}^n \left| \frac{\hat{F} \left(x_i\right) - F \left(x_i\right)}{F \left(x_i\right)} \right|
\end{equation}
where $\{x_i\}$ is the set of observations which make up the background sampling distribution, $n$ is the number of elements of $\{x_i\}$, $\hat{F} \left(x\right)$ is the empirical CDF associated with $\{ x_i \}$, and $F \left(x\right)$ is the theoretical (no veto) or fit (veto) CDF obtained from \eqref{cdf_dEexcess}. For the no veto simulation we find
\begin{equation}
\text{MPE}_\text{no\hspace{1mm}veto} = 0.0125
\end{equation}
indicating that the theoretical PDF/CDF obtained by numerically integrating \eqref{pdf_dEexcess}/\eqref{cdf_dEexcess} with the same parameters used to generate the simulated Gaussian noise accurately describes the simulated detector background. The MPE for the veto simulation is
\begin{equation}
\text{MPE}_\text{veto} = 0.0196
\end{equation}
indicating that the fit of the theoretical model to the simulated detector background with a veto step is also accurate. In addition, we observe that the fit parameters $(n_\text{veto},\tilde{n}_\text{veto}) = (493285,10^{-22}\, \text{Hz}^{-1/2} )$ for the veto case are within $0.4\%$ of the theoretical values $(n_\text{no\hspace{1mm}veto},\tilde{n}_\text{no\hspace{1mm}veto}) = (491520,10^{-22} \, \text{Hz}^{-1/2})$ for the no veto case, demonstrating that the influence of the veto step on the detector background distribution is small.

\subsection{Sensitivity} \label{sec: gwn_sensitivity}

The sensitivity is determined by injecting waveforms with given $h_\text{rss}$ into the background described in Sec.~\ref{sec: det_bg}. Though in general we would expect a GW signal from a magnetar to be elliptically polarized \cite{GWsourcereview}, for simplicity's sake we have used sine-Gaussian waveforms
\begin{equation} \label{SGinj}
h \left( t \right) = A \sin \left( 2 \pi f_c \hspace{0.5mm} t + \phi \right) e^{- \left( \frac{t-t_0}{\tau} \right)^2}
\end{equation}
linearly polarized along the plus-direction, and parameterized by the quality factor $Q = \sqrt{2} \pi \tau f_c \gtrsim 1000$, where $A$ is the peak amplitude, $f_c$ is the central frequency, $\tau$ is the $1/e$ decay time, $\phi$ is the phase of the sinusoid relative to the gaussian envelope, and $t_0$ is the time of the waveform peak. We choose $f_c = 92.5 \,$Hz, corresponding to the QPO at $92.5\,$Hz, and choose values of $t_0$ and $Q$ such that all but a negligible fraction of the injected $h_\text{rss}$ is contained in the interval $\left[ t_\text{center} \hspace{0.5mm} , \hspace{0.5mm} t_\text{center} + \Delta t \right]$. Plus-polarized injections with a given value of $h_\text{rss}$ are constructed by first generating a waveform $h \left( t \right)$ with functional form \eqref{SGinj} and $A = 1$, calculating the energy of this waveform, and then rescaling so that the total energy is given by $F_+^2 h_\text{rss}^2$. The injection waveform satisfies $0 \le E_{h\left(t\right)} \le h_\text{rss}^2$ and $E_{h\left(t\right)}=h_\text{rss}^2$ in the idealized case $F_+ = 1$.

\begin{figure}[h]
\vspace{8mm}
\centering
\includegraphics[width=3.4in]{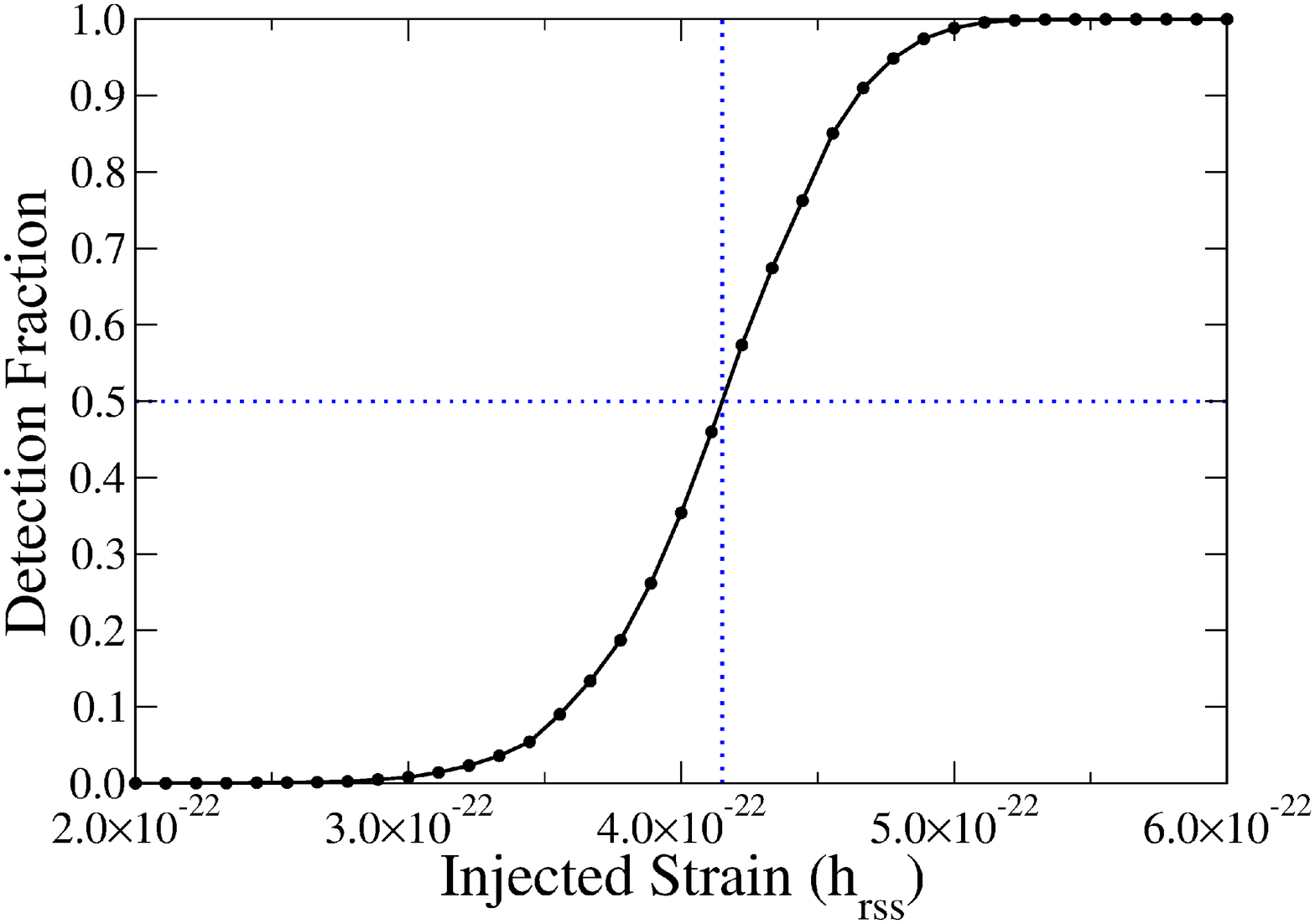}
\caption{A simulation of the sensitivity for Gaussian white noise with strain spectral noise density $\tilde{n} = 10^{-22} \, \text{Hz}^{-1/2}$. The sensitivity is defined to be the minimum injected $h_\text{rss}$ such that the foreground and background distributions are distinguishable via the two-sample K-S test with $5\sigma$ significance $50\%$ of the time. We consider a background sampling distribution with 20,000 samples and a trigger distribution with 108 samples, and use linearly-polarized sine-Gaussian injections. Each injection is performed 100 times assuming ideal detector-source orientation to calculate the detection percentage. \label{fig: sens_results}}
\end{figure}

\begin{figure}[h]
\centering
\includegraphics[width=3.4in]{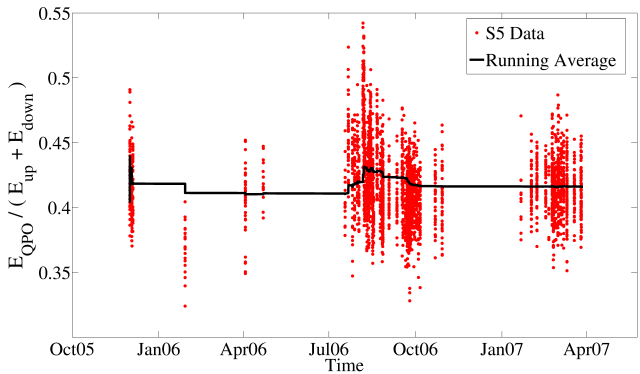}
\caption{$\hat{w}$, defined by \eqref{w^hat}, calculated using H1 data for a series of background times distributed throughout the S5 run up to April $2007$. The running average is computed at time $t$ by averaging all measurements occurring prior to $t$. \label{fig: S5_SF_plot}}
\end{figure}

A foreground sampling distribution is constructed by making one such injection into each of 108 simulated, Gaussian noise segments, where the value 108 is obtained by counting the number of electromagnetic SGR 1806$-$20 triggers observed during S5 for which enough science-quality H1-L1 coincident data is available to compute $\Delta E^\text{excess}$ with $\Delta t = 30\,$s \footnote{For $\Delta t = 300 \,$s there are 82 such triggers.}. The simulated background and foreground distributions are then compared using the two-sample K-S test. This procedure is performed 100 times for each choice of $h_\text{rss}$ to estimate the percentage of simulation trials in which we can claim detection with $5\sigma$ significance \footnote{This corresponds to a K-S test p-value of $p = 2.7 \times 10^{-7}$.}. We iterate by interpolating the resulting injected $h_\text{rss}$ versus detection percentage curve to find the sensitivity, defined to be the minimum injected $h_\text{rss}$ such that we can claim detection 50\% of the time. In Fig.~\ref{fig: sens_results} we examine the sensitivity for a Gaussian noise simulation of the LIGO S5 data set containing 108 triggers. For simplicity we ignore the detector orientation and set $F_+ = F_\times = 1$ for all triggers. As the average value for $F_+$ and $F_\times$ is usually around $0.6$, this simplification underestimates the sensitivity by an average factor of $\sim 1.67$. \\
 \\
We find that the sensitivity with negligible uncertainty for the simulated data set, with trigger times corresponding to those astrophysically detected by satellites, is
\begin{equation} \label{eqn: sens_result}
h_\text{rss}^\text{sens} = 4.15 \times 10^{-22} \, \text{Hz}^{-1/2}.
\end{equation}
Assuming a detector-source distance of $14 \,$kpc for SGR 1806$-$20 \cite{SGR1806-20dist} this corresponds to an isotropic GW emission at the source \eqref{eqn: E_GW} of
\begin{equation}
E_\text{GW}^\text{iso} = 1.10 \times 10^{46} \, \text{erg}
\end{equation}

\section{Application to Off-Source S5 Data} \label{sec: data}

In Sec.~\ref{sec: sim} we demonstrated that \eqref{pdf_dEexcess} accurately describes the background modeled through simulated, Gaussian data. Here we perform a similar analysis using the data from LIGO's fifth science run (S5) for the 4-km Hanford detector (H1) \cite{LIGO}. We first analyze the effects of incorporating the data quality measures described in Sec.~\ref{subsec: DQ} to account for deviations from Gaussianity due to noise contamination in the LIGO data stream. Then we calculate the background $\Delta E^\text{excess}$ distribution of H1 during S5. Finally, we discuss how to extend the algorithm to perform a multi-detector study utilizing the full LIGO network, and optimize the method for SGR 1806$-$20 triggers observed by the H1-L1 network during S5. We again choose $f_\text{QPO} = 92.5 \,$Hz, $\Delta t = 30 \,$s, and $\Delta f = 10 \,$Hz, based on observations of the SGR 1806$-$20 hyperflare event \cite{hyperflare}.

\subsection{S5 Data Quality} \label{sec: S5_DQ}

In order to characterize the weighting factor $w$ we calculate the energy in each of the three channels --- $E_\text{QPO}$, $E_\text{up}$, and $E_\text{down}$ --- in the time interval $[ t_\text{center} - \Delta t, \hspace{0.5mm} t_\text{center} + \Delta t]$ and compute the sampling statistic
\begin{equation} \label{w^hat}
\hat{w} = \frac{E_\text{QPO}}{E_\text{up} + E_\text{down}}.
\end{equation}
In figure 5 this measurement is performed for a series of times $\{ t_\text{center} \}$ distributed throughout the S5 run up to April $2007$. \\
\\
We observe that, on average, $\hat{w} < 0.5$, consistent with the shape of the S5 LIGO noise floor \cite{LIGO}. In addition, the large fluctuations in $\hat{w}$ from sample to sample indicate the need to determine the weighting factor in the excess energy \eqref{Eexcess} locally for each measurement. In the remainder of the paper all $\Delta E^\text{excess}$ measurements are performed using the scheme for determining $w$ described in Sec.~\ref{subsec: DQ}.

\begin{figure}[h]
\vspace{8mm}
\centering
\includegraphics[width=3.4in]{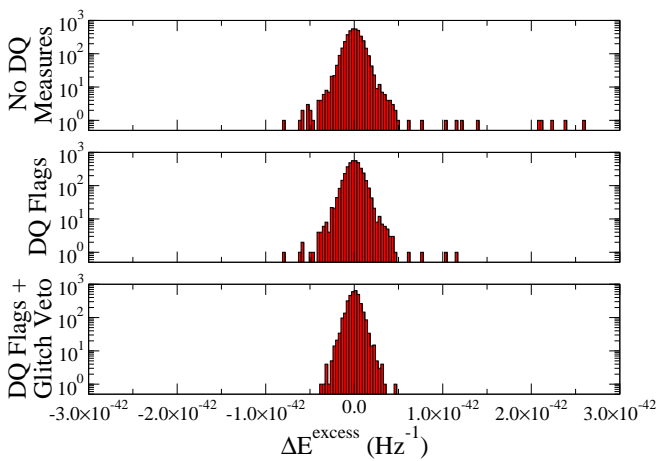}
\caption{Veto methodology and resulting H1 detector background using S5 data. In the top figure no data quality measures are applied. In the middle figure data marked with category 1 or category 2 DQ flags is removed. In the bottom figure category 1 and 2 flags are applied as well as a transient glitch rejection step (see Sec.~\ref{subsec: DQ}) using $10\,$ms tiles and a $4\sigma$ veto threshold. \label{fig: DQ_measures}}
\end{figure}

We investigate the influence of the data quality measures on the shape of the H1 background $\Delta E^\text{excess}$ distribution by performing three trials, corresponding to three veto methodologies. In the first no data quality measures are applied. In the second we exclude data marked with category 1 or category 2 flags. Finally, in the third we incorporate data quality flags as well as a transient glitch rejection step using $10 \,$ms tiles and a $4\sigma$ veto threshold. The resulting distributions are summarized in Fig.~\ref{fig: DQ_measures}.
\\
We conclude that although the inclusion of DQ flags alone goes a long way towards successfully removing the loudest outliers, the glitch veto step is also necessary to fully account for deviations from Gaussianity and to establish strong agreement between the distribution derived from real data and the mathematical model/simulations based on gaussian statistics.

\subsection{H1 Detector Background} \label{sec: S5_bg}

\begin{figure}[h]
\vspace{8mm}
\centering
\includegraphics[width=3.4in]{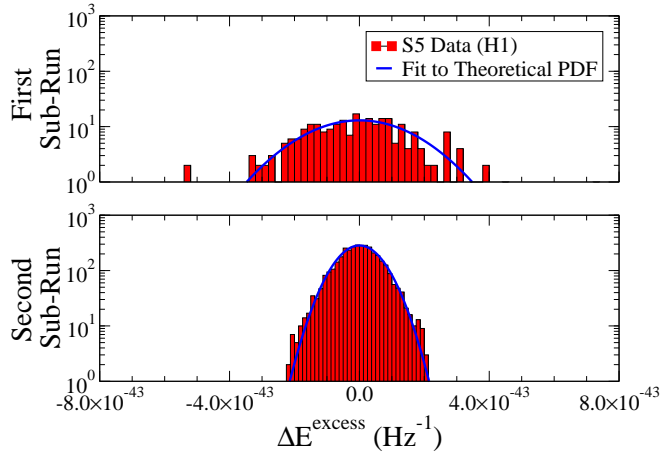}
\caption{H1 detector background (histogram) for each of the sub-runs November 2005 -- June 2006 and June 2006 -- October 2007 using category 1 and category 2 DQ flags as well as a glitch rejection step with $10\,$ms tiles and a $4\sigma$ veto threshold (see Sec.~\ref{subsec: DQ}). For the production of the lower plot, only data before April 2007 was used. The solid curves are fits of the theoretical PDF \eqref{pdf_dEexcess}. The fit parameters are obtained using nonlinear least squares regression to fit the theoretical CDF \eqref{cdf_dEexcess} to the empirical CDF of the background distribution. \label{fig: H1_BG}}
\end{figure}

We find, after incorporating all of the data quality procedures described in Sec.~\ref{subsec: DQ}, that the resulting $\Delta E^\text{excess}$ distribution is accurately described by the theoretical PDF \eqref{pdf_dEexcess} if we partition the S5 run into two, independent sub-runs: the first lasts from the beginning of S5 in November 2005 to June 2006, and the second lasts from June 2006 to the end of S5 in October 2007. The division in June 2006 corresponds to a commissioning break during which the H1 detector received a number of  upgrades. As a result, the detector configurations in each of the sub-runs are considerably different, and, in the context of our analysis, should be treated as distinctly different detector systems. Fig.~\ref{fig: H1_BG} shows the background distributions obtained for each of the two sub-runs. \\
 \\
Given the agreement between the resulting background distributions and the theoretical model \eqref{pdf_dEexcess}, we conclude that our data quality methodology is sufficient to account for non-Gaussianity arising from noise contamination in the S5 data.

\subsection{Extension to Multi-Detector Networks}

Extending the concept to encompass of the full LIGO network requires us to consider how to pool information recorded by multiple instruments into a single analysis, as well as the orientation of each detector at the times of the observed triggers. We extend to a multi-detector network by first performing the single-detector analysis described in Sec.~\ref{sec: pipeline} for each, independent detector. This gives a set of p-values representing the significance with which the two-sample K-S test can distinguish between the background and foreground distributions in the data stream of each detector. Assuming a network of $N$ independent detectors, Fisher's method computes the test statistic
\begin{equation}
X^2 = - 2 \sum\limits_{i = 1}^N \ln p_i
\end{equation}
which follows a $\chi^2$-distribution with $2 N$ degrees of freedom \cite{FishersMethod}. We claim detection for the detector network if $X^2$ is sufficiently large that we can reject the null hypothesis with $5 \sigma$ significance. We choose to explore Fisher's method over a coherent method for two main reasons: (i) Fisher's method does not require us to assume any relationship between the waveforms observed in different detectors beyond the existence of excess energy in the QPO band, and (ii) a significant fraction of triggers will only be recorded by one detector because of downtime, and coherent methods are not applicable for these triggers.

Likewise, we can use Fisher's method to define the sensitivity for the detector network. We obtain a set of antenna factors for a given detector by calculating the antenna response function at the time of each trigger in the Dominant Polarization Frame (DPF) of that detector \footnote{Since we are computing the energy of the injected waveform, which is polarization independent, we are free to pick a different polarization frame for each detector.}. For a single detector, this is the frame in which the detector response is entirely in the plus direction ($F_\times \equiv 0$) \cite{DPF}. We then get a multi-detector sensitivity result by first fixing a value of $h_\text{rss}$, and then performing the single-detector sensitivity analysis described in Sec.~\ref{sec: gwn_sensitivity} for each detector in the network individually, weighting injections with the appropriate set of antenna factors. We then combine the single-detector results using Fisher's method, and define the sensitivity of the network to be the minimum injected $h_\text{rss}$ such that we can reject the null hypothesis with $5 \sigma$ significance in $50\%$ of trials. In the remainder of the paper we will consider the H1-L1 network, where L1 is the 4-km Livingston detector.

\begin{figure}[h]
\begin{tabular}{cc}
\centerline{\includegraphics[width=3.4in]{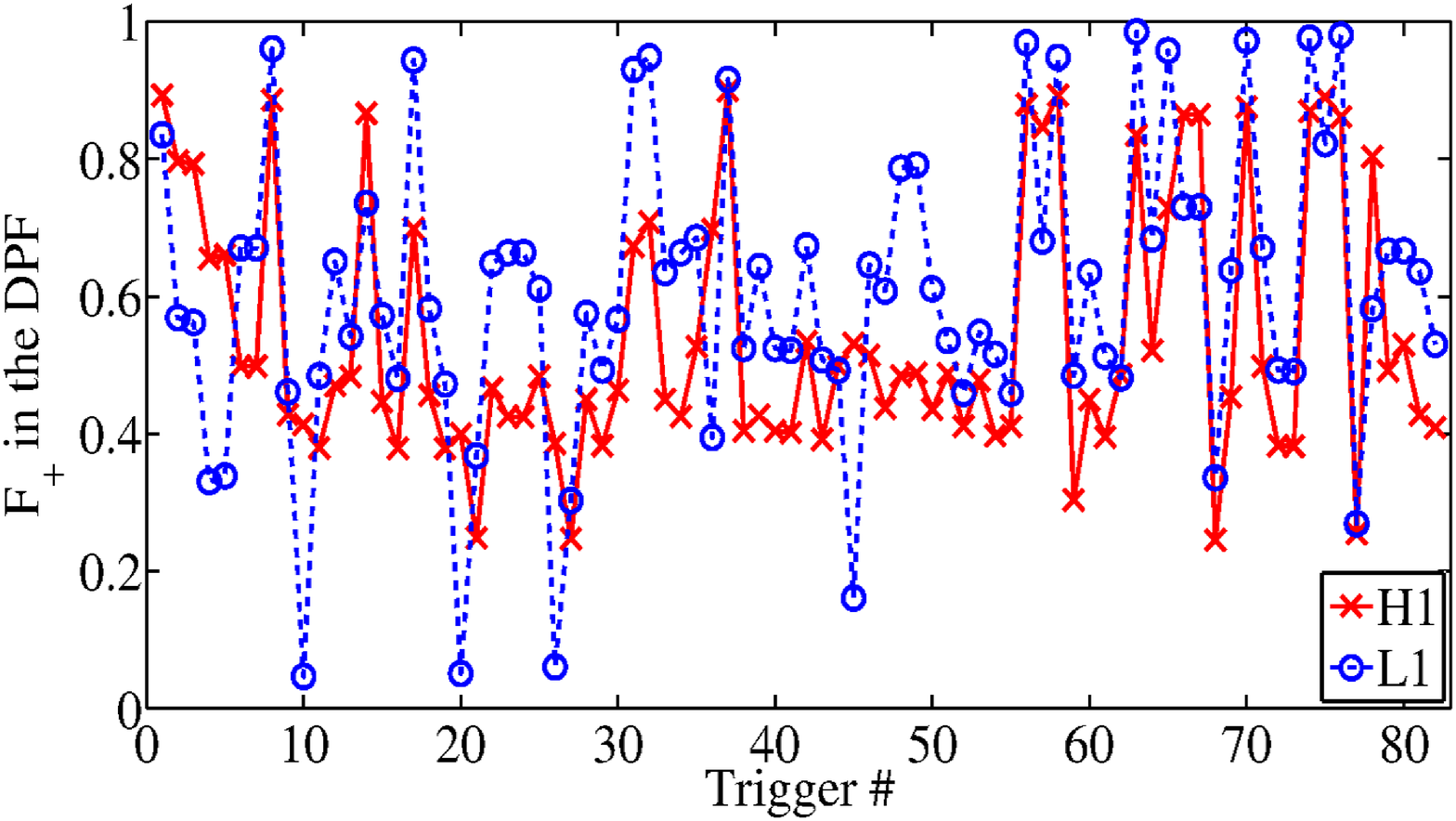}} \\
\centerline{\includegraphics[width=3.4in]{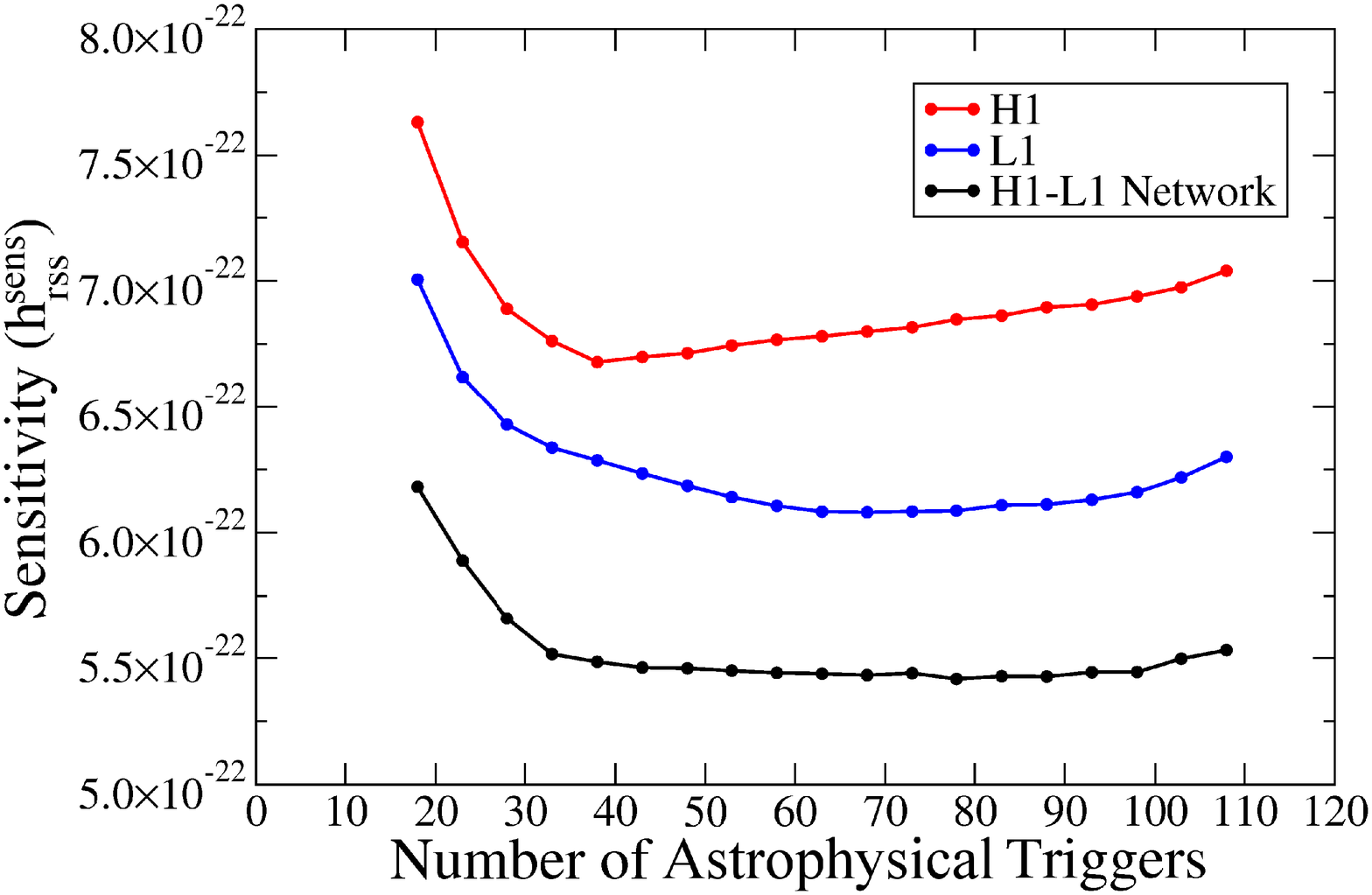}}
\end{tabular}
\caption{The upper plot shows the $F_+$ values for H1 and L1 in their respective Dominant Polarization Frames (DPF-s, see \cite{DPF}) calculated for the direction of the SGR 1806$-$20 source at the times of the astrophysical triggers observed during S5. In the lower plot we compute the sensitivity for Gaussian noise with strain spectral noise density $\tilde n = 10^{-22} \, \text{Hz}^{-1/2}$ as a function of the number of triggers included in the analysis with the highest corresponding $F_+$ values. In each case we choose the number of these triggers such as to maximize sensitivity (i.e. minimize $h_\text{rss}^\text{sens}$). \label{fig: AF}}
\end{figure}

In addition, we must consider the fact that the detector network is not necessarily equally sensitive to all triggers observed from a single source since the relative positions of the detectors and source change with time. Fig.~\ref{fig: AF}a shows the quantity $F_+$ for each trigger and for H1 and L1 separately. Since the algorithm becomes more sensitive as the number of triggers increases, but is less sensitive if the detector is poorly aligned for some of the triggers, we expect that there is an optimal subset of the full list of triggers for which the sensitivity of a multi trigger approach is maximized. We perform this optimization by repeating the sensitivity analysis using simulated, Gaussian noise described in Sec.~\ref{sec: gwn_sensitivity} for H1 and L1 individually, weighting the injection waveforms with the appropriate antenna factors, and then combining the results using Fisher's method to get a sensitivity for the hypothetical H1-L1 network. We then find the worst-aligned member (smallest value of $F_+$) of the combined list of H1 and L1 triggers, exclude this trigger for both detectors, and compute the sensitivity again. In Fig.~\ref{fig: AF}b we plot the sensitivity as a function of the number of triggers, obtained by iterating this process.
 \\
We conclude that the sensitivity of the algorithm to GW emissions from SGR 1806$-$20 is maximized for the S5 H1-L1 data set if we restrict only to the 78 triggers for which the detectors were best aligned with the source. The same procedure can easily be repeated to tune the algorithm for other data sets or sources. Furthermore, after repeating the simulation of Sec.~\ref{sec: gwn_sensitivity} with the detector orientation taken into account we find that the single detector sensitivities are
\begin{eqnarray}
\hspace{-20mm} \left( h_\text{rss}^\text{sens} \right)_\text{H1} &=& 6.68 \times 10^{-22} \, \text{Hz}^{-1/2} \\
\left( h_\text{rss}^\text{sens} \right)_\text{L1} &=& 6.08 \times 10^{-22} \, \text{Hz}^{-1/2}
\end{eqnarray}
and the sensitivity for the H1-L1 network obtained using Fisher's method is
\begin{equation}
\left( h_\text{rss}^\text{sens} \right)_\text{H1-L1} = 5.42 \times 10^{-22} \, \text{Hz}^{-1/2}.
\end{equation}
Again assuming a detector-source distance of $14\,$kpc for SGR 1806$-$20, the corresponding isotropic GW emission energies are
\begin{eqnarray}
\left( E_\text{GW}^\text{iso} \right)_\text{H1} &= 2.84 \times 10^{46} \, \text{erg} \\
\left( E_\text{GW}^\text{iso} \right)_\text{L1} &= 2.35 \times 10^{46} \, \text{erg} \\
\left( E_\text{GW}^\text{iso} \right)_\text{H1-L1} &= 1.87 \times 10^{46} \, \text{erg}
\end{eqnarray}

\section{Conclusion}

We have investigated aspects of a multi-trigger data analysis concept targeting the detection of long-duration narrow-band gravitational wave transients. Potential astrophysical motivators include quasi-periodic oscillations of magnetars that are already observed in the electromagnetic spectrum following soft gamma repeater hype-flare events. This concept significantly extends a single-trigger method introduced by Matone and M\'{a}rka \cite{OldMethods} both in data handling sophistication and comprehensive use of available external information. The method is based on an excess-energy statistic targeting long-duration, narrow-band signals, and incorporates a number of data quality measures to account for nonstationarity in the data stream of a physical detector over long periods of time. The detection threshold and sensitivity are both defined in terms of the two-sample Kolmogorov-Smirnov test as a viable example.

Analytical results for the PDF and CDF of the detector background distribution enabled us to demonstrate that these expressions accurately describe the output of the pipeline applied to (i) Gaussian white noise and (ii) off-source LIGO S5 data from the 4-km Hanford detector after reasonable data quality cuts. In turn, they showed that the justified data quality cuts are efficient in removing the major contamination from the data. It is important to note that a useful side product of the method might be the ability to identify distinct running condition of the gravitational wave detectors (for a similar work in this topic, see \cite{STAMP_DetChar}).

We also demonstrated how to combine the data streams of multiple, independent detectors into a single study on an alternative non-coherent way, and illustrated how to optimize the algorithm for a future search involving the S5 Hanford-Livingston data set. Simulating the optimal trigger list and using Gaussian white noise with strain spectral noise density $\tilde{n} = 10^{-22} \, \text{Hz}^{-1/2}$ resulted in an upper bound on the total energy emitted in isotropic gravitational radiation by SGR 1806$-$20 in the 30 seconds following a flare of
\begin{equation}
\left( E_\text{GW}^\text{iso} \right)_\text{H1-L1} = 1.87 \times 10^{46} \, \text{erg} \times \left( \frac{d}{14 \, \text{kpc}} \right)^2,
\end{equation}
where $d$ is the detector-source distance.

Due to the uncertainties of currently available model predictions, throughout our study we assumed that (i) the central frequency, bandwidth, and time scale of a gravitational wave signal associated with an SGR QPO event can directly be taken from the electromagnetic observations of the QPO signal, (ii) the duration of the gravitational wave signals are the same for every QPO event of the same SGR source, and (iii) the duration of all the gravitational wave signals associated with the astrophysical triggers from an SGR source can be approximated with the duration of the longest flare event among the triggers (e.g. in case of SGR 1806$-$20, it is the 27 December 2004 hyperflare event). These assumptions however might turn out to be too optimistic as theoretical model predictions become more sophisticated in the future. Our method can be applied to any chosen central frequency, and can be generalized to a wide scale of bandwidths. However, a more detailed study is necessary to see the consequences of having different durations for the gravitational wave signals emitted during the different flare events of the same SGR source. We leave this study to be part of a future work.

\section{Acknowledgments}

This paper was reviewed by the LIGO Scientific Collaboration under LIGO Document $\mathrm{P1300011}$. The authors are grateful for the support of the LIGO-Virgo Collaboration, and Columbia University in the City of New York. The authors would like to thank Jolien Creighton, Rutupurna Das, David Fierroz, Raymond Frey, Alyssa Miller, Peter Shawhan, and Eric Thrane for their helpful discussions and contributions. The research was supported by the National Science Foundation under grants PHY-0457528, PHY-0847182, and PHY-0856691, and by Columbia University in the City of New York.

\end{document}